\documentclass[sigconf,nonacm]{acmart}

\usepackage{subfigure}
\usepackage{siunitx}
\usepackage{enumitem}

\usepackage{algorithm}
\usepackage[noend]{algpseudocode}
\let\ReturnInline\Return
\renewcommand{\Return}{\State\ReturnInline}
\algrenewcommand\algorithmicrequire{$\rhd$}
\algrenewcommand\algorithmicensure{$\square$}

\AtBeginDocument{%
  \providecommand\BibTeX{{%
    \normalfont B\kern-0.5em{\scshape i\kern-0.25em b}\kern-0.8em\TeX}}}

\setcopyright{acmcopyright}
\copyrightyear{2018}
\acmYear{2018}
\acmDOI{XXXXXXX.XXXXXXX}

\acmConference[Conference acronym 'XX]{Make sure to enter the correct
  conference title from your rights confirmation emai}{June 03--05,
  2018}{Woodstock, NY}

\newcommand{\ignore}[1]{}

\begin{document}

\title[Memory-Efficient Community Detection on Large Graphs Using Weighted Sketches]{Memory-Efficient Community Detection on Large Graphs \\Using Weighted Sketches}


\author{Subhajit Sahu}
\email{subhajit.sahu@research.iiit.ac.in}
\affiliation{%
  \institution{IIIT Hyderabad}
  \streetaddress{Professor CR Rao Rd, Gachibowli}
  \city{Hyderabad}
  \state{Telangana}
  \country{India}
  \postcode{500032}
}


\settopmatter{printfolios=true}

\begin{abstract}
Community detection in graphs identifies groups of nodes with denser connections within the groups than between them, and while existing studies often focus on optimizing detection performance, memory constraints become critical when processing large graphs on shared-memory systems. We recently proposed efficient implementations of the Louvain, Leiden, and Label Propagation Algorithms (LPA) for community detection. However, these incur significant memory overhead from the use of collision-free per-thread hashtables. To address this, we introduce memory-efficient alternatives using weighted Misra-Gries (MG) sketches, which replace the per-thread hashtables, and reduce memory demands in Louvain, Leiden, and LPA implementations --- while incurring only a minor quality drop (up to $1\%$) and moderate runtime penalties. We believe that these approaches, though slightly slower, are well-suited for parallel processing and could outperform current memory-intensive techniques on systems with many threads.
\end{abstract}



\maketitle

\section{Introduction}
\label{sec:introduction}
Research on graph-structured data has seen rapid growth, driven by the capacity of graphs to represent complex, real-world interactions and capture intricate relationships between entities. At the core of this field is community detection, a technique that divides graphs into tightly connected subgroups or communities, thereby revealing the natural structure within the data. Community detection finds applications across a wide range of areas, including examining epidemic-prone group dynamics \cite{salathe2010dynamics}, studying zoonotic eco-epidemiology \cite{desvars2024one}, detecting diseases like lung cancer \cite{bechtel2005lung}, categorizing tumors via genomic data \cite{haq2016community}, aiding therapeutic discovery \cite{ma2019comparative, udrescu2020uncovering}, mapping healthcare areas \cite{wang2021network}, analyzing retail patterns \cite{verhetsel2022regional}, identifying transportation trends \cite{chen2023deciphering}, unsupervised part-of-speech tagging \cite{das2011unsupervised}, partitioning graphs for machine learning \cite{bai2024leiden}, automating microservice decomposition \cite{cao2022implementation}, sectionalizing power systems \cite{aziz2023novel}, characterizing polarized information ecosystems \cite{uyheng2021mainstream}, identifying hidden social network groups \cite{blekanov2021detection, la2022information}, detecting disinformation on Telegram \cite{la2021uncovering}, investigating restored Twitter accounts \cite{kapoor2021ll}, mapping multi-step cyberattacks \cite{zang2023attack}, detecting blockchain attacks \cite{erfan2023community}, studying cyber resilience \cite{chernikova2022cyber}, analyzing human brain networks \cite{bullmore2009complex, he2010graph}, and understanding metabolic network evolution \cite{pfeiffer2005evolution, kim2009centralized}. Community detection is also used for addressing other graph related problems, such as, finding connected components \cite{stergiou2018shortcutting}, graph partitioning \cite{meyerhenke2017parallel, slota2020scalable}\ignore{\cite{slota2014pulp, wang2014partition, meyerhenke2014partitioning, meyerhenke2016partitioning, bae2020label, akhremtsev2020high, zhang2020multilevel}}\ignore{, hypergraph partitioning \cite{henne2015label, gottesburen2021scalable}}, vertex reordering and graph compression \cite{boldi2011layered}, and graph coarsening \cite{valejo2020coarsening}.

Community detection is challenging due to the lack of prior knowledge about the number of communities and their size distribution, a problem that has led to the development of various heuristic methods for identifying communities \cite{com-blondel08, com-gregory10, com-raghavan07, com-newman16, com-ghoshal19}. A commonly used metric for assessing the quality of detected communities is the modularity score, introduced by Newman et al. \cite{com-newman04}. The Louvain method, introduced by Blondel et al. \cite{com-blondel08}, is a widely used community detection algorithm \cite{com-lancichinetti09} that applies a two-phase approach consisting of an iterative local-moving phase and an aggregation phase to optimize the modularity metric across multiple passes. However, Traag et al. \cite{com-traag19} found that the Louvain method can yield poorly connected and even internally disconnected communities. They proposed the Leiden algorithm, which introduces an additional refinement phase to address these shortcomings, enabling the algorithm to better detect well-connected communities \cite{com-traag19}. The Label Propagation Algorithm (LPA) is another method that outperforms the above algorithms in terms of speed and scalability, but yields communities with lower modularity scores. However, it has been observed to achieve high Normalized Mutual Information (NMI) score compared to the ground truth \cite{peng2014accelerating}.

Given the importance of the community detection problem, a number of existing studies have aimed at improving the performance of the above algorithms using various algorithmic optimizations \cite{com-rotta11, com-waltman13, com-gach14, com-traag15, com-lu15, com-ozaki16, com-naim17, com-halappanavar17, com-ghosh18, com-traag19, com-shi21, com-xing14, com-berahmand18, com-sattari18, com-you20, com-liu20}\ignore{\cite{com-ryu16, com-zhang21, com-you22, com-aldabobi22}} and parallelization techniques \cite{com-cheong13, com-wickramaarachchi14, com-lu15, com-naim17, com-fazlali17, com-halappanavar17, com-ghosh18, com-bhowmik19, com-shi21, com-bhowmick22, staudt2015engineering, soman2011fast, kuzmin2015parallelizing, traag2023large}\ignore{\cite{com-zeng15, com-que15, com-zeitz17, com-gheibi20}}. Additionally, significant effort has gone into developing efficient parallel implementations for multicore CPUs \cite{staudt2015engineering, staudt2016networkit, com-fazlali17, com-halappanavar17, qie2022isolate, huparleiden}, GPUs \cite{com-naim17, kang2023cugraph}, CPU-GPU hybrids \cite{com-bhowmik19, com-mohammadi20}, multi-GPUs \cite{com-cheong13, kang2023cugraph, chou2022batched, com-gawande22}, and multi-node systems --- CPU only \cite{com-ghosh18, ghosh2018scalable, sattar2022scalable, huparleiden} / CPU-GPU hybrids \cite{com-bhowmick22}. However, these studies focus primarily on reducing the runtime of the algorithms\ignore{, rather than its memory usage}. As network sizes grow, the memory footprint becomes a critical concern, particularly when processing large graphs on shared-memory systems. Recently, we proposed some of the most efficient implementations of Louvain \cite{sahu2023gvelouvain}, Leiden \cite{sahu2023gveleiden}, and LPA \cite{sahu2023gvelpa}. These implementations have a space complexity of $O(T|V| + |E|)$, where $|V|$ is the number of vertices, $|E|$ is the number of edges, and $T$ is the number of threads used. As a result, they also face similar memory constraints.

In this work, we present a method based on the Misra-Gries heavy hitters algorithm \cite{misra1982finding} to significantly reduce memory usage in our Louvain,\footnote{\url{https://github.com/puzzlef/louvain-lowmem-communities-openmp}} Leiden,\footnote{\url{https://github.com/puzzlef/leiden-lowmem-communities-openmp}} and LPA\footnote{\url{https://github.com/puzzlef/rak-lowmem-communities-openmp}} implementations, with minimal impact on community quality. While this approach introduces some runtime overhead, it is more parallelizabile, and by current trends, may eventually outperform existing memory-intensive methods.

\section{Related work}
\label{sec:related}
We now review studies on community detection in the edge streaming setting, where graphs are presented as sequences of edges that must be processed in a single pass. The proposed algorithms in these works are designed to minimize both runtime and memory usage, allowing for efficient processing of graph streams. Hollocou et al. \cite{hollocou2017linear, hollocou2017streaming} introduce the Streaming Community Detection Algorithm (SCoDA), which maintains only two to three integers per node. This approach leverages the observation that randomly chosen edges are more likely to connect nodes within the same community than between different communities. Wang et al. \cite{wang2023streaming} focus on identifying the local community around specific query nodes in graph streams. Their online, single-pass algorithm samples the neighborhood around query nodes, using an approximate conductance metric to extract the target community from the sampled subgraph. Liakos et al. \cite{liakos2017coeus, liakos2020rapid} discuss approaches for detecting communities in graph streams by expanding seed-node sets as edges arrive, without retaining the full graph structure.

Although these streaming algorithms effectively balance runtime and memory efficiency for detecting communities, they are limited by the single-pass constraint. This may reduce the quality of detected communities compared to algorithms that access the entire graph and perform multiple passes over edges to refine communities. In this technical report, we propose techniques to reduce the memory usage of the widely used Louvain, Leiden, and LPA algorithms, with little to no compromise in the quality of communities obtained, as measured by the modularity metric.



\section{Preliminaries}
\label{sec:preliminaries}
Consider an undirected graph $G = (V, E, w)$, where $V$ represents the set of vertices, $E$ is the set of edges, and $w_{ij} = w_{ji}$ denotes the positive weight of each edge $(i, j)$. For unweighted graphs, each edge is assumed to have a weight of one, i.e., $w_{ij} = 1$. For any vertex $i$, the set of its neighbors is given by $J_i = \{ j \mid (i, j) \in E \}$. The weighted degree of vertex $i$ is defined as $K_i = \sum_{j \in J_i} w_{ij}$, summing the weights of all edges incident to $i$. Additionally, the graph has $N = |V|$ vertices and $M = |E|$ edges. The total sum of edge weights across the entire graph is denoted by $m = \frac{1}{2} \sum_{i,j \in V} w_{ij}$.

\subsection{Community detection}
\label{sec:about-communities}

Communities based solely on a network's structure, without external data, are termed intrinsic. These communities are disjoint if each vertex belongs to only one community \cite{com-gregory10}. Disjoint community detection aims to establish a community membership function $C: V \rightarrow \Gamma$, which assigns each vertex $i \in V$ a community ID $c \in \Gamma$, where $\Gamma$ is the set of community IDs. The vertices in community $c$ are represented as $V_c$, and the community of vertex $i$ is denoted $C_i$. The neighbors of vertex $i$ in community $c$ are defined as $J_{i \rightarrow c} = \{ j \ | \ j \in J_i \ \text{and} \ C_j = c \}$. The total edge weight between vertex $i$ and its neighbors in community $c$ is $K_{i \rightarrow c} = \{ w_{ij} \ | \ j \in J_{i \rightarrow c} \}$, while the sum of edge weights within community $c$ is $\sigma_c = \sum_{(i, j) \in E \ \text{and} \ C_i = C_j = c} w_{ij}$, and the total edge weight associated with $c$ is $\Sigma_c = \sum_{(i, j) \in E \ \text{and} \ C_i = c} w_{ij}$.

\subsection{Modularity}
\label{sec:about-modularity}

Modularity is a metric for assessing the quality of communities formed by community detection algorithms, typically based on heuristics \cite{com-newman04}. Its value ranges from $-0.5$ to $1$, with higher values indicating stronger community structures. It measures the difference between the actual and expected fractions of edges within communities if edges were randomly assigned \cite{com-brandes07}. The modularity score, $Q$, for identified communities is calculated using Equation \ref{eq:modularity}. Additionally, the change in modularity --- referred to as delta modularity --- when moving vertex $i$ from community $d$ to community $c$, represented as $\Delta Q_{i: d \rightarrow c}$, is defined by Equation \ref{eq:delta-modularity}.

\begin{equation}
\label{eq:modularity}
  Q
  = \sum_{c \in \Gamma} \left[\frac{\sigma_c}{2m} - \left(\frac{\Sigma_c}{2m}\right)^2\right]
\end{equation}

\begin{align}
\begin{split}
\label{eq:delta-modularity}
  &\Delta Q_{i: d \rightarrow c} = \Delta Q_{i: d \rightarrow i} + \Delta Q_{i: i \rightarrow c} \\
  &= \left[ \frac{\sigma_d - 2K_{i \rightarrow d}}{2m} - \left(\frac{\Sigma_d - K_i}{2m}\right)^2 \right] + \left[ 0 - \left(\frac{K_i}{2m}\right)^2 \right] - \left[ \frac{\sigma_d}{2m} - \left(\frac{\Sigma_d}{2m}\right)^2 \right] \\
  &+ \left[ \frac{\sigma_c + 2K_{i \rightarrow c}}{2m} - \left(\frac{\Sigma_c + K_i}{2m}\right)^2 \right] - \left[ \frac{\sigma_c}{2m} - \left(\frac{\Sigma_c}{2m}\right)^2 \right] - \left[ 0 - \left(\frac{K_i}{2m}\right)^2 \right] \\
  &= \frac{1}{m} (K_{i \rightarrow c} - K_{i \rightarrow d}) - \frac{K_i}{2m^2} (K_i + \Sigma_c - \Sigma_d)
\end{split}
\end{align}


\subsection{Louvain algorithm}
\label{sec:about-louvain}

The Louvain method, proposed by Blondel et al. \cite{com-blondel08}, is a greedy, agglomerative algorithm designed to optimize modularity for detecting high-quality communities within a graph. It operates in two main phases. In the local-moving phase, each vertex $i$ evaluates the potential benefits of joining a neighboring community $C_j$ (where $j \in J_i$) to maximize the increase in modularity $\Delta Q_{i:C_i \rightarrow C_j}$. The second phase involves merging vertices from the same community into super-vertices. These phases repeat until no further modularity gains can be achieved, resulting in a hierarchical structure (dendrogram) where the top level represents the final communities.

\subsection{Leiden algorithm}
\label{sec:about-leiden}

However, Traag et al. \cite{com-traag19} found that the Louvain method frequently generates poorly-connected and even internally-disconnected communities. To address these issues, they developed the Leiden algorithm, which incorporates a refinement phase after the local-moving phase. This phase enables the algorithm to identify communities that are both well-separated and well-connected. During refinement, vertices start with singleton sub-communities, and merge with suitable adjacent sub-communities if no other vertices have joined their own community. This merging occurs once (i.e., for a single iteration) and is randomized, with the likelihood of a vertex joining a neighboring community based on the delta-modularity of the move. The Leiden algorithm has a time complexity of $O(L|E|)$, where $L$ is the number of iterations, and a space complexity of $O(|V| + |E|)$, similar to that of the Louvain method.

\subsection{Label Propagation Algorithm (LPA)}
\label{sec:about-rak}

The Label Propagation Algorithm (LPA), proposed by Raghavan et al. \cite{com-raghavan07}, is a popular diffusion-based algorithm for identifying medium-quality communities in large networks. Its advantages include simplicity, speed, and scalability. In the algorithm, each vertex $i$ begins with a unique label $C_i$ (its community ID), and during each iteration, vertices adopt the label that has the highest interconnecting weight. This iterative process leads to consensus among densely connected vertex groups, converging when at least $1 - \tau$ of vertices (where $\tau$ is the tolerance parameter) retain their community labels. The algorithm has a time complexity of $O(L |E|)$ and a space complexity of $O(|V| + |E|)$, with $L$ representing the number of iterations. While LPA typically yields communities with lower modularity scores compared to the Louvain and Leiden algorithms, it has been observed to achieve high Normalized Mutual Information (NMI) score relative to the ground truth \cite{peng2014accelerating}.


\subsection{Boyer-Moore (BM) majority vote algorithm}
\label{sec:about-bm}

The Boyer-Moore (BM) majority vote algorithm \cite{boyer1991mjrty} efficiently identifies the majority element in a sequence --- defined as an element that appears more than $n/2$ times in a list of $n$ elements. It was developed by Robert S. Boyer and J. Strother Moore, and published in 1981. The algorithm maintains a candidate for the majority element along with a counter for its "votes." Initially, it sets the \textit{candidate} and \textit{count} variables. As it iterates through the list, if the count is $0$, it assigns the current element as the new candidate and sets the count to $1$. If the count is not zero and the current element matches the candidate, it increments the count; if it differs, it decrements the count. By the end of the iteration, the candidate variable holds the potential majority element. This algorithm is efficient, completing in a single pass with $O(n)$ time complexity and using constant space $O(1)$.

\subsection{Misra-Gries (MG) heavy hitters algorithm}
\label{sec:about-mg}

The Misra-Gries (MG) heavy hitters algorithm, introduced in 1982 by Jayadev Misra and David Gries \cite{misra1982finding}, extends the Boyer-Moore majority finding algorithm for the heavy-hitter problem. It aims to find elements that occur more than $\frac{n}{k+1}$ times, where $n$ is the total number of processed elements and $k+1$ is a user-defined threshold. The algorithm initializes an empty set of up to $k$ counters, each tracking a \textit{candidate} element and its approximate \textit{count}. As elements are processed, if an element is already a candidate, its counter is incremented. If it is not a candidate and fewer than $k$ counters are used, a new counter is created with a count of $1$. If all $k$ counters are occupied, the algorithm decrements each counter by $1$. Counters that reach zero, result in the corresponding element being removed from the candidate set. After processing the stream, the remaining candidates are those that likely exceed the $\frac{n}{k}$ threshold, although a verification step may be needed to confirm their exact counts. The decrementing mechanism effectively prunes infrequent elements from the list. The MG algorithm operates with a time complexity of $O(n)$ and a space complexity of $O(k)$, making it suitable for scenarios with limited computational resources.

\section{Approach}
\label{sec:approach}
\begin{figure}[hbtp]
  \centering
  \subfigure{
    \label{fig:about-hashtable--all}
    \includegraphics[width=0.88\linewidth]{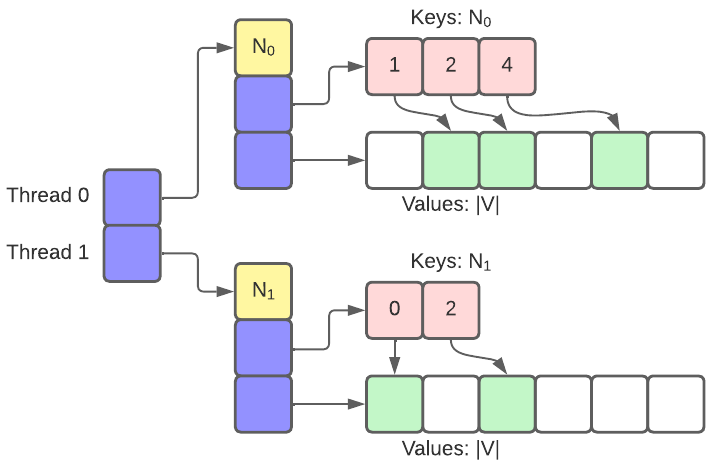}
  } \\[-2ex]
  \caption{Illustration of collision-free, per-thread hashtables (Far-KV) that are well spaced in memory, for two threads. Each hashtable includes a keys list, a values array (of size $|V|$), and a key count ($N_0$ and $N_1$). Values associated with each key are stored and accumulated at the index specified by the key. To prevent false cache sharing, the key counts for each hashtable are independently allocated on the heap.}
  \label{fig:about-hashtable}
\end{figure}

We have recently proposed one of the most efficient multicore implementations of the Louvain, Leiden, and LPA algorithms --- which we refer to as GVE-Louvain \cite{sahu2023gvelouvain}, GVE-Leiden \cite{sahu2023gveleiden}, and GVE-LPA \cite{sahu2023gvelpa}, respectively. These implementations employ collision-free hashtables for each thread, used in the local-moving and aggregation phases of GVE-Louvain, in the local-moving, refinement, and aggregation phases of GVE-Leiden, and in every iteration of GVE-LPA. Each hashtable consists of a keys list, a values array (of size $|V|$, the number of vertices in the graph), and a key count. Each value is stored or accumulated at the index matching its key. To avoid false cache sharing, the key count of each hashtable is updated independently and allocated separately on the heap. This significantly reduces or eliminates conditional branching and minimizes the number of instructions required to insert or accumulate entries. An illustration of these hashtables is shown in Figure \ref{fig:about-hashtable}.

However, these hashtables incur a substantial memory overhead, with space usage ranging from $8T|V|$ to $16T|V|$ bytes (depending on the number of populated entries, with values stored as 64-bit floating-point numbers and keys as 64-bit integers), i.e., they have a space complexity of $O(T|V|)$. Here, $T$ denotes the number of threads used for community detection. For instance, processing a graph with $100$ million vertices using $64$ threads would require between $51.2$ GB and $102.4$ GB of memory for the hashtables alone. With larger graphs, this memory demand can escalate rapidly, This motivates us to explore strategies for reducing the memory footprint of the hashtables, even if it entails some trade-offs in performance.

Our first attempt at reducing the memory usage focuses on scaling down the size of the hashtable’s values array by a factor of $3000\times$. Our findings indicate that this has only a minimal effect on the quality of generated communities. We elaborate on this method in Section \ref{sec:small-hashtables}. We now move on to present our main method proposed in this paper, which further reduces the memory required for each hashtable to just $0.5$ KB or less, irrespective of graph size.

\begin{figure*}[hbtp]
  \centering
  \subfigure{
    \label{fig:about-move--all}
    \includegraphics[width=0.78\linewidth]{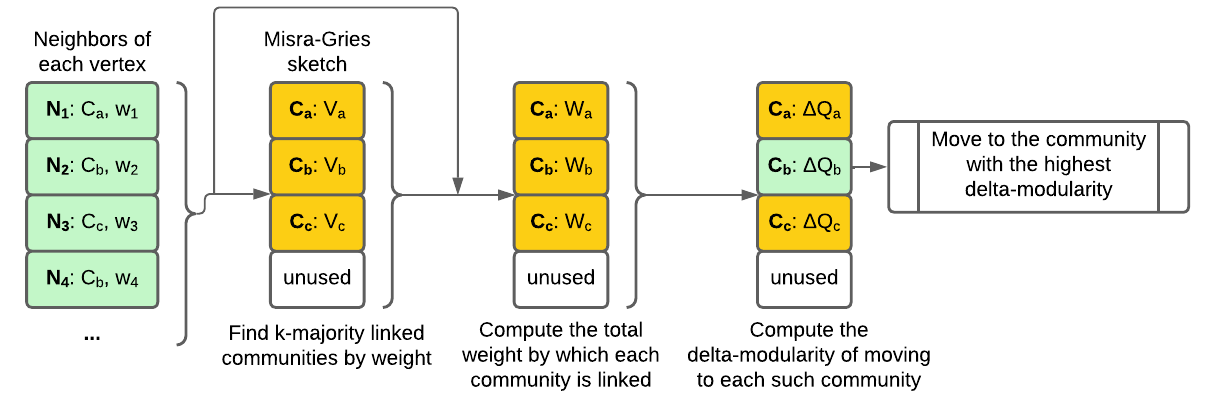}
  } \\[-2ex]
  \caption{Illustration of our modification to the local-moving phase of the Louvain algorithm, the local-moving and refinement phases of the Leiden algorithm, and the iterations of LPA. Here, $N_1$, $N_2$, $N_3$, $N_4$ denote the neighbors of a vertex $i$ in the graph, with associated edge weights to $i$ of $w_1$, $w_2$, $w_3$, $w_4$, respectively, while $C_a$, $C_b$, $C_c$ denote the community memberships of the neighbors. This is used to populate a weighted Misra-Gries (MG) sketch with $k$ slots, where $V_a$, $V_b$, and $V_c$ in the MG sketch represent the accumulated weights for communities $C_a$, $C_b$, and $C_c$, respectively. Once populated, a second pass is made over vertex $i$’s edges to compute the total linking weights $W_a$, $W_b$, and $W_c$ for the majority candidate communities $C_a$, $C_b$, and $C_c$ --- which is then used to compute the delta-modularities $\Delta Q_a$, $\Delta Q_b$, $\Delta Q_c$ of moving $i$ to the candidate communities $C_a$, $C_b$, $C_c$, respectively. Finally, the community $C_b$ yielding the highest positive modularity gain $\Delta Q_b$ is selected as $i$’s new community.}
  \label{fig:about-move}
\end{figure*}

\subsection{For Louvain algorithm}

In every iteration of the local-moving phase in the Louvain algorithm, each vertex $i \in V$ in the input graph $G$ examines its neighboring vertices $J_i$ to identify the neighboring community $c^*$ that would provide the largest gain in modularity if joined. To achieve this, the algorithm begins by accumulating the edge weights of each neighbor $j \in J_i$ of vertex $i$ into a hashtable, where the keys correspond to each neighbor's community membership $C_j$. This yields the total edge weight $K_{i \rightarrow c}$ between $i$ and each neighboring community $c \in \Gamma_i$, where $\Gamma_i = \{C_j \mid j \in J_i\}$. Using the data stored in the hashtable, the algorithm then calculates the modularity gain $\Delta Q_{i \rightarrow c}$ for each potential move of $i$ to a neighboring community $c$, following Equation \ref{eq:delta-modularity}, and selects the community $c^*$ that offers the highest $\Delta Q$. If such a $c^*$ is identified, vertex $i$ is moved to this community, along with updating the total edge weight of the previous and current community of $i$. However, if all potential moves yield a negative $\Delta Q$, vertex $i$ remains in its current community.

We now turn to minimizing the memory footprint of the \textit{local-moving phase} of Louvain algorithm. Our approach is based on the idea that a fully populated map of neighboring communities $c \in \Gamma_i$ for each vertex $i$ along with the associated linking weights $K_{i \rightarrow c}$ is not necessary. Instead, we can work with a ``sketch" of this information, focusing only on the most significant neighboring communities --- specifically, those with a linking weight greater than $\frac{K_i}{k+1}$, where $K_i$ represents the weighted degree of vertex $i$, and $k$ is a user-defined parameter. The reasoning is that the community with the highest delta-modularity, $c^*$, is likely be among these $k$-majority candidate communities. To this end, we use a weighted version of the Misra-Gries (MG) heavy-hitter algorithm \cite{misra1982finding}, with $k$ slots. Here, rather than counting the occurrences of each linked community $c \in \Gamma_i$ for a vertex $i \in V$, we accumulate the weights of edges connecting $i$ to its neighboring communities. Given that we utilize $k$ slots, once the edge weights for all neighbors $j \in J_i$ (grouped by their community memberships $C_j$) are accumulated, the MG algorithm will have identified up to $k$-majority \textit{candidate} communities. As with the unweighted MG algorithm, not all neighboring communities $c \in \Gamma_i$ will necessarily have a linking weight above $K_i / k$, so remaining entries may include non-majority communities, or may be empty if $|\Gamma_i| < k$. We then perform a second pass over vertex $i$'s edges in order to determine the total linking weight $K_{i \rightarrow c}$ between $i$ and each of these $k$-majority communities, calculate the delta-modularity $\Delta Q_{i \rightarrow c}$ for moving vertex $i$ into each of them, and select the community with the highest delta-modularity. This chosen community, $c^\#$,\ignore{then} becomes the new community for $i$\ignore{in the local-moving phase}. It is important to note that $c^\#$ may differ from $c^*$; however, with an appropriately chosen $k$, they are expected to align in most cases. Figure \ref{fig:about-move} provides an illustration of the above process.

\begin{figure*}[hbtp]
  \centering
  \subfigure{
    \label{fig:about-aggregate--all}
    \includegraphics[width=0.63\linewidth]{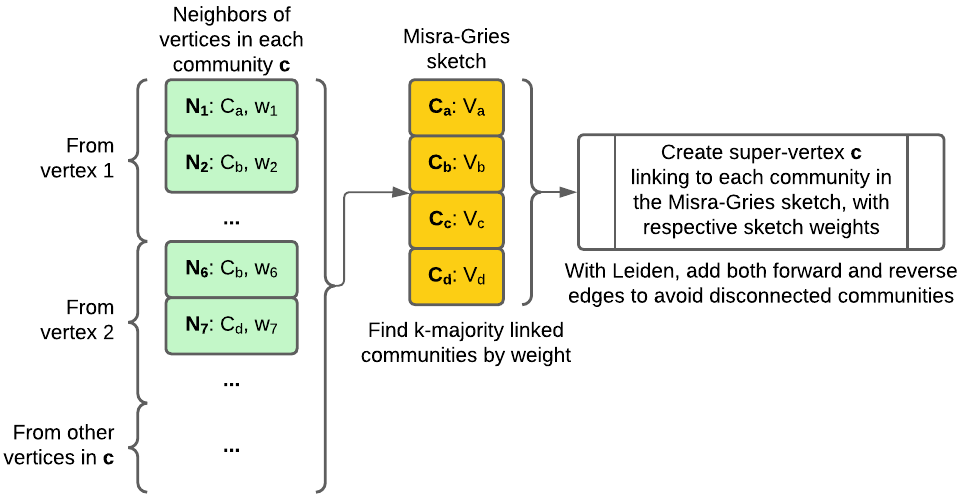}
  } \\[-2ex]
  \caption{Illustration of our modification of the aggregation phase in the Louvain and Leiden algorithms. Here, $N_1$ and $N_2$ are neighbors of a vertex in community $c$, while $N_6$ and $N_7$ are neighbors of another vertex in the same community --- with associated edge weights $w_1$, $w_2$, $w_6$, and $w_7$ to their respective source vertex, and with associated community memberships $C_a$, $C_b$, $C_c$, and $C_d$. These are used to populate a weighted Misra-Gries (MG) sketch with $k$ slots, where $V_a$, $V_b$, $V_c$, and $V_d$ in the MG sketch represent the cumulative weights for communities $C_a$, $C_b$, $C_c$, and $C_d$, respectively. Once populated, a super-vertex $c$ is created in the aggregated graph, linking to each community $C_a$, $C_b$, $C_c$, and $C_d$ with the corresponding weights $V_a$, $V_b$, $V_c$, and $V_d$. To prevent disconnected communities, our modified Leiden also ensures that the aggregated graph remains undirected.}
  \label{fig:about-aggregate}
\end{figure*}

In the aggregation phase of the Louvain algorithm, the objective is to simplify the graph $G$ by constructing a new super-vertex graph $G'$ in which each community identified in the local-moving phase is represented as a single vertex (super-vertex). This process begins by identifying the vertices that belong to each community. For every community $c \in \Gamma$, the algorithm then iterates through all linked communities $d$ (each with an associated edge weight $w$) connected to each vertex $i$ in community $c$, adding these to a hashtable. Once the hashtable is filled with all communities and their corresponding weights associated with community $c$, these can then be added as edges to super-vertex $c$ in the super-vertex graph $G'$. The process is repeated until all communities in $G$ are represented as super-vertices in the super-vertex graph $G'$. The local-moving phase of the Louvain algorithm is then executed on the newly created super-vertex graph $G'$ in the subsequent iteration of the algorithm, continuing until convergence is achieved.

We now discuss how we reduce the memory usage during the \textit{aggregation phase} of Louvain algorithm. Similar to the local-moving phase, we employ the weighted version of MG algorithm to accumulate and identify $k$-majority neighboring communities for each community derived from the local-moving phase. To achieve this, we iterate over each vertex $i \in V$ where $C_i = c$ for each community $c \in \Gamma$ in the graph $G$, processing the neighbors $J_i$ of all such vertices, one after the other. Following this, we add the identified $k$ majority communities, along with their accumulated values, as edges to the super-vertex $c$ in the super-vertex graph $G'$. This process is repeated until each community in $G$ is represented as a super-vertex in $G'$. In contrast to the local-moving phase, however, we do not conduct a second pass over the edges of each vertex within the communities --- to determine the precise cross-community edge weights to the $k$-majority communities from each community $c \in \Gamma$ --- as this would significantly increase computational cost. Our results indicate that this does not badly affect the quality of generated communities. We illustrate the above process in Figure \ref{fig:about-aggregate}.

\begin{figure*}[!hbt]
  \centering
  \subfigure{
    \label{fig:optlou--runtime}
    \includegraphics[width=0.48\linewidth]{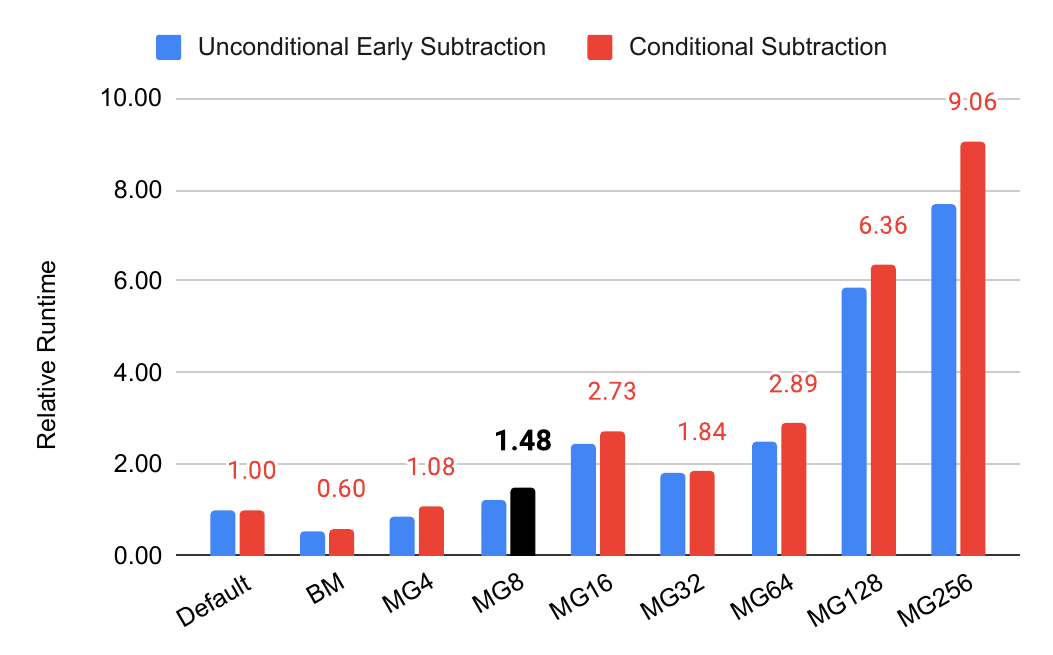}
  }
  \subfigure{
    \label{fig:optlou--modularity}
    \includegraphics[width=0.48\linewidth]{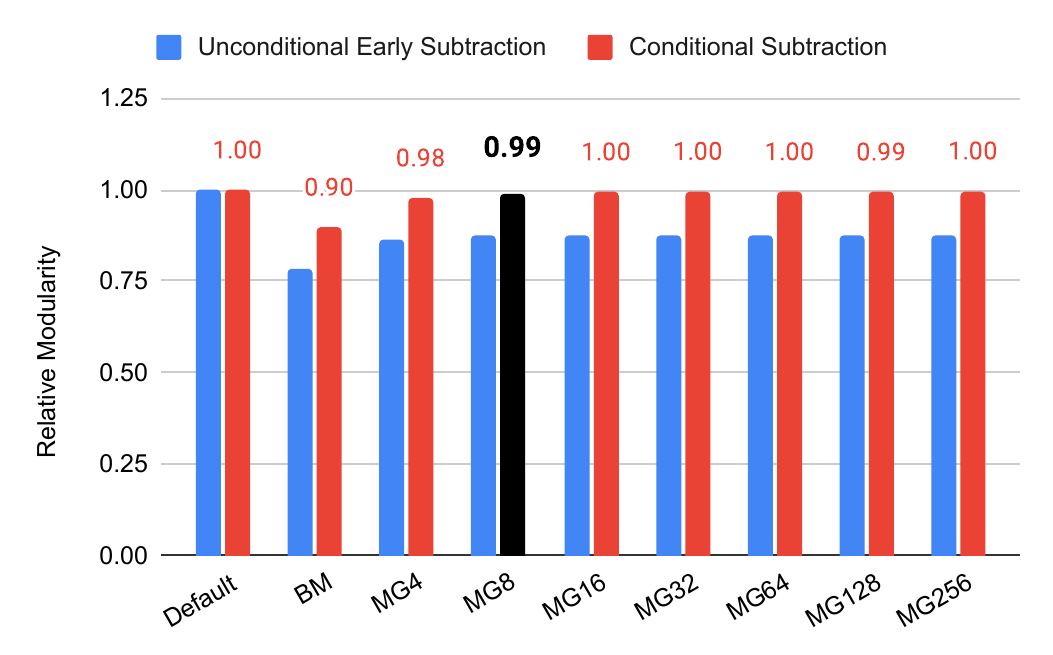}
  } \\[-2ex]
  \caption{Relative Runtime and Relative Modularity of communities obtained from Default, Boyer-Moore (BM), and Misra-Gries (MG) based Louvain, with slot counts ranging from $4$ to $256$ in powers of $2$. Two variations of each MG-based Louvain are compared: one that unconditionally subtracts values from all non-matching keys before inserting a new key-value pair into the Misra-Gries sketch, and another that performs conditional subtraction only after a failed insertion attempt. Although Default Louvain has only one method, it is shown as two variations for simplicity. The most suitable approach is highlighted\ignore{in the figure}.}
  \label{fig:optlou}
\end{figure*}

We now aim to determine a suitable value for $k$, representing the number of slots --- to be used during the local-moving and aggregation phases of the Louvain algorithm --- with minimal or no compromise in community quality, as compared to GVE-Louvain \cite{sahu2023gvelouvain} (which we refer to as the default implementation of the Louvain algorithm)\ignore{ which utilizes a keys list and a full-sized values array as previously detailed}. Specifically, we define a $99\%$ threshold for community quality, compared to GVE-Louvain. To accomplish this, we vary the number of slots in the Misra-Gries (MG) sketch from $4$ to $256$ in powers of $2$. We evaluate two variations for each MG-based Louvain approach: (1) an \textit{unconditional subtraction} of values from all non-matching keys before inserting a new key-value pair (if the new key is not already in the sketch), and (2) a \textit{conditional subtraction}, applied only if the insertion of the new key-value pair fails (due to the absence of a free slot). Additionally, we explore minimizing the memory footprint of the local-moving phase using a weighted version of the Boyer-Moore (BM) majority vote algorithm \cite{boyer1991mjrty}. This is effectively a minimal instance of the weighted MG algorithm with $k = 1$, tracking only a single majority candidate community. However, we do not apply the BM algorithm for the aggregation phase, as it would yield degenerate, chain-like communities. Instead, when using weighted BM for the local-moving phase, we opt for the weighted MG algorithm with $k = 4$ for aggregation. We conduct these experiments on large real-world graphs (listed in Table \ref{tab:dataset}), ensuring the graphs are undirected and weighted with a default weight of $1$. Figure \ref{fig:optlou} illustrates the relative runtime and modularity of communities returned by the Default, BM-based, and MG-based Louvain algorithms with varying slot numbers ($4$ to $256$, in powers of $2$). Note that Default Louvain only has a single approach, despite being presented in the figure as having two different approaches. As the figure shows, our modified Louvain algorithm, incorporating the Misra-Gries (MG) sketch with $k = 8$ slots, achieves a favorable balance between runtime and modularity, while substantially lowering memory demand for the per-thread hashtables used in the algorithm --- with the runtime of our modified Louvain algorithm being only $1.48\times$ slower and results in just a $1\%$ decrease in community quality (measured by modularity).

\begin{figure*}[!hbt]
  \centering
  \subfigure{
    \label{fig:optlei--runtime}
    \includegraphics[width=0.48\linewidth]{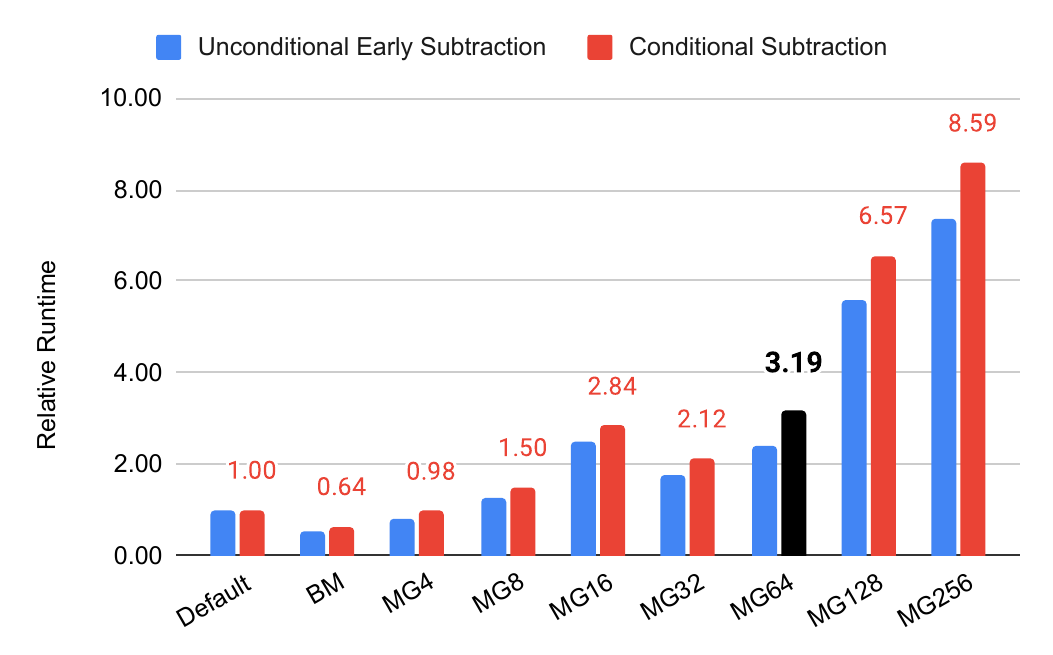}
  }
  \subfigure{
    \label{fig:optlei--modularity}
    \includegraphics[width=0.48\linewidth]{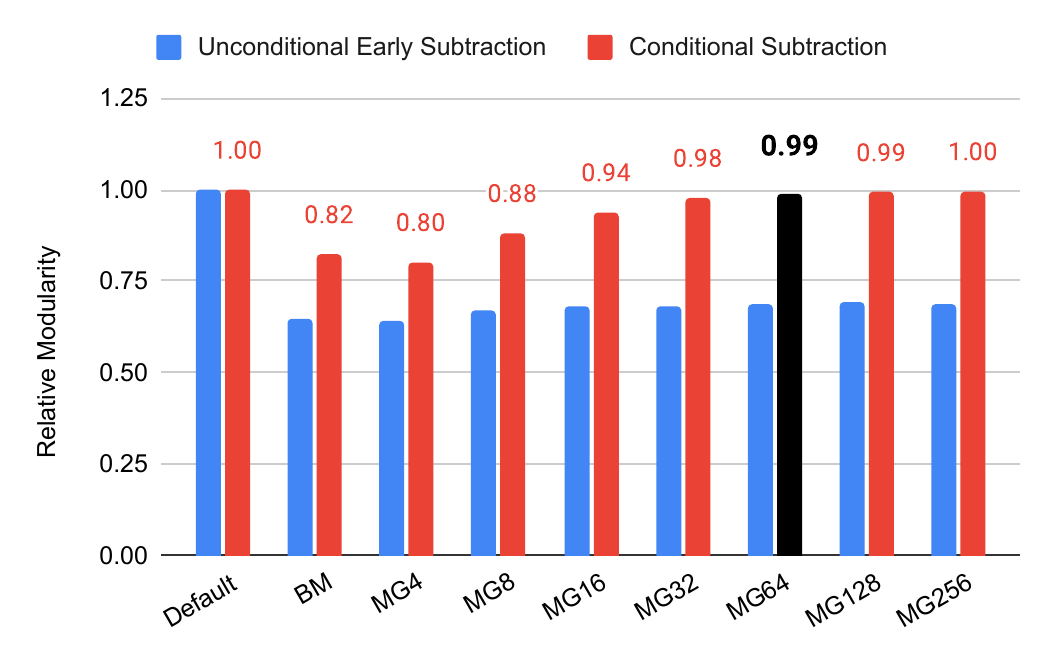}
  } \\[-2ex]
  \caption{Relative Runtime and Relative Modularity of communities from Default, Boyer-Moore (BM), and Misra-Gries (MG) based Leiden, with the number of slots ranging from $4$ to $256$ in powers of $2$. Two variations of each MG-based Leiden are compared: one that unconditionally subtracts values from all non-matching keys before inserting a new key-value pair into the MG sketch, and another that performs conditional subtraction only after a failed insertion attempt. Although Default Leiden is a single method, it is shown as two variations for simplicity. The most suitable approach is highlighted in the figure.}
  \label{fig:optlei}
\end{figure*}

\subsection{For Leiden algorithm}

The Leiden algorithm is similar to the Louvain algorithm, but it adds a refinement phase between the local-moving and aggregation phases. In this phase, each vertex $i \in V$ in the input graph $G$ evaluates its neighboring vertices $J_i$ to find the community $c^*$ that would yield the greatest modularity gain if joined, similar to the local-moving phase. However, unlike the local-moving phase, a vertex moves to a neighboring community only if no other vertices have joined its current community. Additionally, each vertex can only move to sub-communities within its own community as identified during the local-moving phase, and it is processed only once. The refinement phase of the Leiden algorithm makes similar use of per-thread hashtables as the local-moving phase.

To minimize the memory footprint of Leiden algorithm, we use a process similar to that of the Louvain algorithm. Specifically, the same approach is applied during the local-moving and refinement phases of the Leiden algorithm as in the Louvain algorithm's local-moving phase, as illustrated in Figure \ref{fig:about-move}. The aggregation phase of the Leiden algorithm follows a similar procedure as the Louvain algorithm's aggregation phase, as illustrated in Figure \ref{fig:about-aggregate}. However, we ensure that the super-vertex graph is undirected. This step is necessary because our low-memory aggregation scans do not guarantee that the super-vertex graph will be undirected, even if the input graph is undirected --- \textit{leading to disconnected communities}. To address this, we atomically add both forward and reverse edges during each CSR edge insertion, while ensuring that neither edge is already present in the super-vertex graph.

We next determine an appropriate value for $k$ (the number of slots) to use in the local-moving, refinement, and aggregation phases of our MG-based Leiden algorithm. As earlier, we set a quality threshold of $99\%$ relative to GVE-Leiden \cite{sahu2023gveleiden} (hereafter referred to as the default Leiden algorithm) and vary the slots count in the MG sketch from $4$ to $256$. We analyze two variations for each MG-based Leiden method: (1) \textit{unconditional subtraction}, where values are subtracted from all non-matching keys before inserting a new key-value pair (if the new key is not already in the sketch); and (2) \textit{conditional subtraction}, where values are only subtracted if the new key insertion fails because no free slots are available. Additionally, we examine a weighted variant of the BM algorithm, applying the BM algorithm in the local-moving and refinement phases, but employing the weighted MG algorithm with $k = 4$ for the aggregation phase. Experiments are conducted on large, real-world graphs (see Table \ref{tab:dataset}), after ensuring that each graph is both undirected and weighted, with a weight of $1$ for each edge in the graph.

Figure \ref{fig:optlei} presents the relative runtime and modularity of communities returned by the Default, BM-based, and MG-based Leiden algorithms, with slot counts ranging from $4$ to $256$ (in powers of $2$). As the results reveals, our modified Leiden algorithm, which integrates the Misra-Gries (MG) sketch with $k = 64$ slots, runs $3.19\times$ slower than the default but delivers only a $1\%$ reduction in community quality, while significantly reducing memory footprint. The increased slot requirement for our modified Leiden algorithm is likely due to the refinement phase, where sub-communities are small, and each vertex has many possible sub-communities to join, with no choice being significantly better than the others --- thus requiring a higher value of $k$ in order to ensure that the linked weight $K_{i \rightarrow c^*}$ for the optimal community $c^*$ falls within $K_i/k$.

\begin{figure*}[!hbt]
  \centering
  \subfigure{
    \label{fig:optrak--runtime}
    \includegraphics[width=0.48\linewidth]{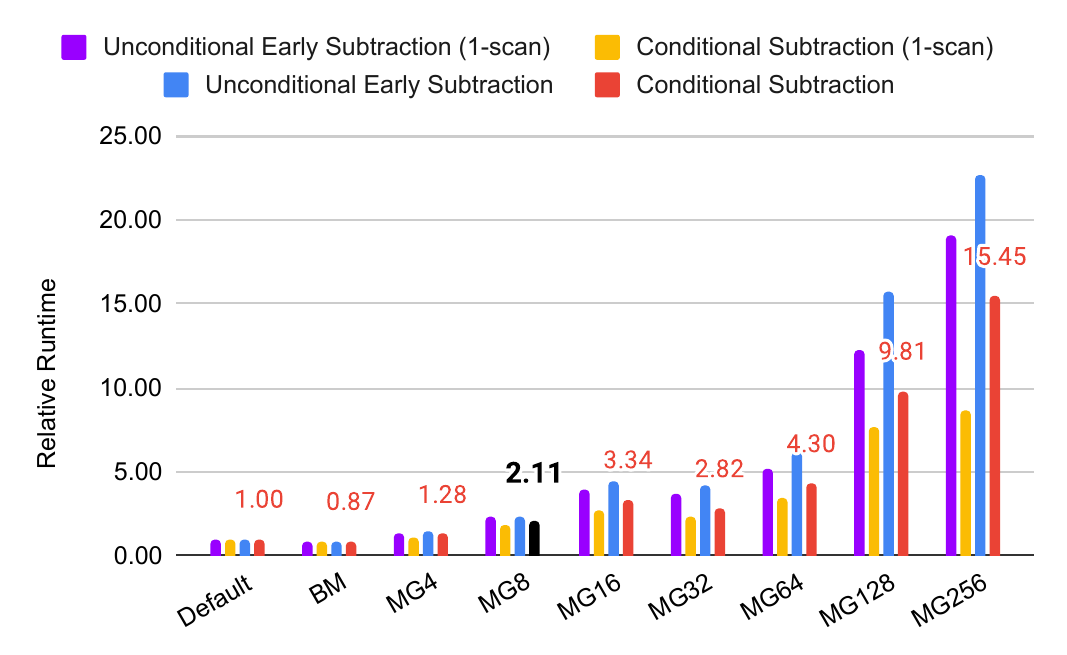}
  }
  \subfigure{
    \label{fig:optrak--modularity}
    \includegraphics[width=0.48\linewidth]{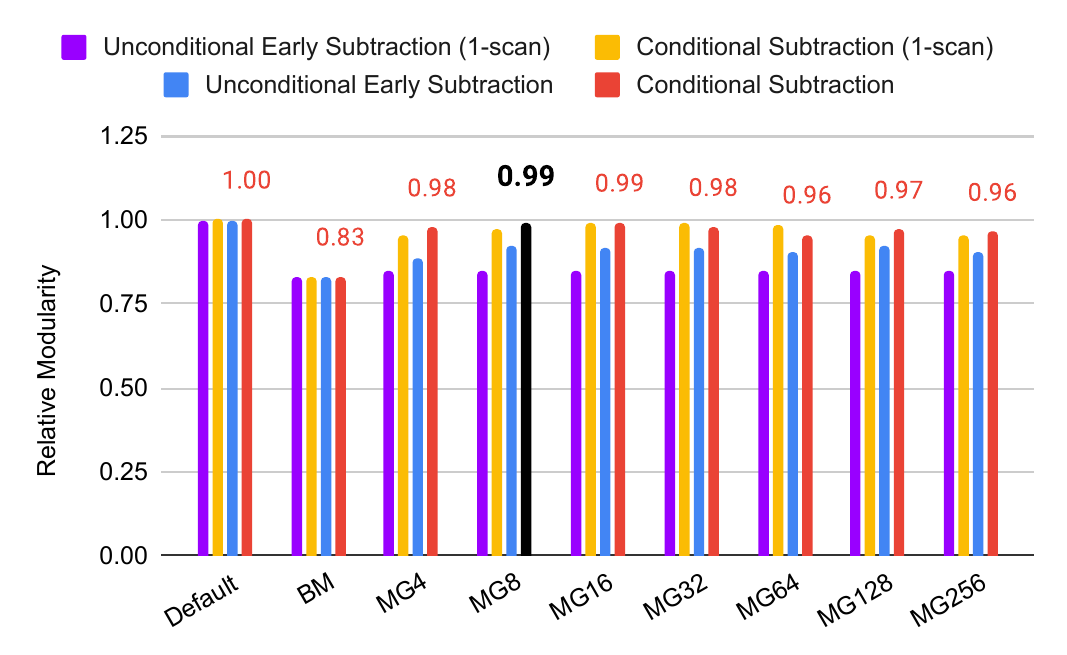}
  } \\[-2ex]
  \caption{Relative Runtime and Relative Modularity of communities from Default, Boyer-Moore (BM), and Misra-Gries (MG) based LPA, with slot counts ranging from $4$ to $256$\ignore{ in powers of $2$}. We compare two variations of the MG-based LPA: one that unconditionally subtracts values from all non-matching keys before inserting a new key-value pair into the MG sketch, and another that performs conditional subtraction only after a failed insertion attempt. Additionally, a single scan approach (\textit{$1$-scan}) is tested for both variations, which omits the calculation of total linking weight $K_{i \rightarrow c}$ between the current vertex $i$ and its $k$-majority communities, instead selecting the majority community based solely on the highest accumulated value in the MG sketch. Although Default LPA is a single method, it is displayed as two variations for clarity. The most effective approach is highlighted.}
  \label{fig:optrak}
\end{figure*}


\subsection{For Label Propagation Algorithm (LPA)}

In every iteration of LPA, each vertex $i \in V$ in the graph $G$ iterates over its neighbors $J_i$, excluding itself, and calculates the total edge weight $K_{ i \rightarrow c}$ for each unique label $c \in \Gamma_i$ present among its neighboring vertices. These weights are stored in a per-thread hashtable. The label $c^*$ with the highest total edge weight $K_{ i \rightarrow c^*}$ is then chosen from the hashtable as the new label for vertex $i$.

To minimize the memory usage of LPA, we use an approach akin to the local-moving phase of our MG-based Louvain, as depicted in Figure \ref{fig:about-move} --- but do not compute the delta-modularity of moving the current vertex $i$ to each $k$-majority candidate community. Instead, we just perform a second pass over the edges of $i$ to determine the total edge weight $K_{i \rightarrow c}$ between $i$ and each of its $k$-majority communities, and then select the majority community $c^\#$ with the highest linking weight as the new label for vertex $i$. As mentioned previously, $c^\#$ may not always match $c^*$; nonetheless, with a suitably chosen $k$, they are expected to align in most instances.

Next, we determine an appropriate value for $k$ (the number of slots) to use in the iterations of our MG-based LPA. As before, we establish a quality threshold of $99\%$ relative to GVE-LPA \cite{sahu2023gvelpa}, referred to here as the default LPA, and vary the slot count in the MG sketch from $4$ to $256$. We analyze two variations of the MG-based LPA: (1) \textit{unconditional subtraction}, where values are subtracted from all non-matching keys before inserting a new key-value pair (provided the new key is not already in the sketch); and (2) \textit{conditional subtraction}, where values are only subtracted if the insertion of the new key fails due to a lack of available slots. For both variations, we also test a single scan approach (\textit{$1$-scan}), which bypasses the calculation of the total linking weight $K_{i \rightarrow c}$ between the current vertex $i$ and each of its $k$-majority communities, and instead select the majority community $c^\#$ based solely on the highest accumulated value in the MG sketch --- helping us to reduce runtime, potentially without sacrificing the quality of the resulting communities. Furthermore, we explore a weighted variant of the BM algorithm, which serves as a minimal instance of the weighted MG algorithm with $k = 1$, by focusing on a single majority candidate community. Our experiments are conducted on large, real-world graphs (Table \ref{tab:dataset}), while ensuring that each graph is undirected and unit-weighted by default.

Figure \ref{fig:optrak} shows the relative runtime and modularity of communities identified by Default, BM-based, and MG-based LPA, with $k$ varying from $4$ to $256$ (in powers of $2$). Results indicate that our modified LPA, which utilizes MG sketch with $k = 8$ slots and includes a second pass, is $2.11\times$ slower than Default LPA, but only incurs a $1\%$ decrease in community quality, while significantly decreasing memory usage. The single scan approach ($1$-scan) does not yield better performance as it does not effectively select the best label\ignore{as the double scan approach does}.

\subsection{Space complexity}

In our MG-based implementation of Louvain, Leiden, and LPA algorithms, each thread's memory requirement for its MG sketch is minimal, needing only $k \leq 64$ slots (with $k = 8$ for MG-based Louvain and LPA, and $k = 64$ for MG-based Leiden). As a result, the space complexity of our algorithm is $O(|V|+|E|)$, or $O(|V|)$ if we disregard the memory for storing the input graph. The pseudocode for populating the weighted version of MG sketch, which we leverage in our MG-based Louvain, Leiden, and LPA, is provided in Algorithm \ref{alg:populate}, with further explanation in Section \ref{sec:populate}.

\section{Evaluation}
\label{sec:evaluation}
\subsection{Experimental setup}
\label{sec:setup}

\subsubsection{System}
\label{sec:system}

We use a server that is powered by two Intel Xeon Gold 6226R processors, each offering $16$ cores with a clock speed of $2.90$ GHz. Each core includes a $1$ MB L1 cache, a $16$ MB L2 cache, and a shared $22$ MB L3 cache. The system is equipped with $376$ GB of RAM and operates on CentOS Stream 8.

\subsubsection{Configuration}
\label{sec:configuration}

We represent vertex IDs using 32-bit unsigned integers and edge weights with 32-bit floating-point numbers. For aggregating floating-point data, we switch to 64-bit floating-point precision. The primary parameters for our MG-based Louvain, Leiden, and LPA algorithms mirror those used in GVE-Louvain \cite{sahu2023gvelouvain}, GVE-Leiden \cite{sahu2023gveleiden}, and GVE-LPA \cite{sahu2023gvelpa}. For MG-based Louvain and LPA, we set $k = 8$, while for MG-based Leiden we use $k = 64$. Here, $k$ defines the number of slots in the MG sketch, which serves as an alternative to traditional per-thread hash tables (Figure \ref{fig:about-hashtable}). All implementations are executed with 64 threads and compiled using GCC 13.2 with OpenMP 5.1, using the \texttt{-O3} and \texttt{-mavx} optimization flags. Figure \ref{fig:about-mg8code} presents the C++ source code for our MG8 kernel (using $k=8$ slots), alongside its compiled machine code translation.

\subsubsection{Dataset}
\label{sec:dataset}

We conduct experiments on $13$ large real-world graphs from the SuiteSparse Matrix Collection, as listed in Table \ref{tab:dataset}. These graphs span from $3.07$ million to $214$ million vertices and from $25.4$ million to $3.80$ billion edges. In our experiments, we ensure that all edges are undirected, by inserting missing reverse edges, and weighted, with a default weight of $1$. We did not use publicly available real-world weighted graphs in this study due to their small size. However, our parallel algorithms are capable of handling weighted graphs without modification.


\begin{table}[hbtp]
  \centering
  \caption{List of $13$ graphs retrieved from the SuiteSparse Matrix Collection \cite{suite19} (with directed graphs indicated by $*$). Here, $|V|$ denotes the number of vertices, $|E|$ denotes the number of edges (after making the graph undirected by adding reverse edges), and $|\Gamma|$ denotes the number of communities obtained with \textit{Static Leiden} algorithm \cite{sahu2024fast}.}
  \label{tab:dataset}
  \begin{tabular}{|c||c|c|c|}
    \toprule
    \textbf{Graph} &
    \textbf{\textbf{$|V|$}} &
    \textbf{\textbf{$|E|$}} &
    \textbf{\textbf{$|\Gamma|$}} \\
    \midrule
    \multicolumn{4}{|c|}{\textbf{Web Graphs (LAW)}} \\ \hline
    indochina-2004$^*$ & 7.41M & 341M & 2.68K \\ \hline
    uk-2002$^*$ & 18.5M & 567M & 41.8K \\ \hline
    arabic-2005$^*$ & 22.7M & 1.21B & 2.92K \\ \hline
    uk-2005$^*$ & 39.5M & 1.73B & 18.2K \\ \hline
    webbase-2001$^*$ & 118M & 1.89B & 2.94M \\ \hline
    it-2004$^*$ & 41.3M & 2.19B & 4.05K \\ \hline
    sk-2005$^*$ & 50.6M & 3.80B & 2.67K \\ \hline
    \multicolumn{4}{|c|}{\textbf{Social Networks (SNAP)}} \\ \hline
    com-LiveJournal & 4.00M & 69.4M & 3.09K \\ \hline
    com-Orkut & 3.07M & 234M & 36 \\ \hline
    \multicolumn{4}{|c|}{\textbf{Road Networks (DIMACS10)}} \\ \hline
    asia\_osm & 12.0M & 25.4M & 2.70K \\ \hline
    europe\_osm & 50.9M & 108M & 6.13K \\ \hline
    \multicolumn{4}{|c|}{\textbf{Protein k-mer Graphs (GenBank)}} \\ \hline
    kmer\_A2a & 171M & 361M & 21.1K \\ \hline
    kmer\_V1r & 214M & 465M & 10.5K \\ \hline
  \bottomrule
  \end{tabular}
\end{table}

\subsubsection{Measurement}
\label{sec:measurement}

We evaluate the runtime of each method encompassing all phases of the algorithm. To reduce the impact of noise in our experiments, we follow standard practice of repeating each experiment multiple times. Further, we assume the total edge weight of the graphs is known. Given that modularity maximization is an NP-hard problem, and that existing polynomial-time algorithms are heuristic, we assess the optimality of our algorithms by comparing their convergence to the modularity score achieved by the default implementation of respective algorithms, i.e., GVE-Louvain, GVE-Leiden, and GVE-LPA. Lastly, our MG-based Leiden algorithm does not generate internally disconnected communities, similar to GVE-Leiden, so we exclude this detail from our figures.

\subsection{Performance of our MG-based Louvain}
\label{sec:performance-louvain}

\begin{figure*}[hbtp]
  \centering
  \subfigure[Runtime in seconds (logarithmic scale) with \textit{Default}, \textit{weighted Boyer-Moore (BM) based}, and \textit{weighted Misra-Gries with $8$-slots (MG8) based} Louvain]{
    \label{fig:cmplou--runtime}
    \includegraphics[width=0.98\linewidth]{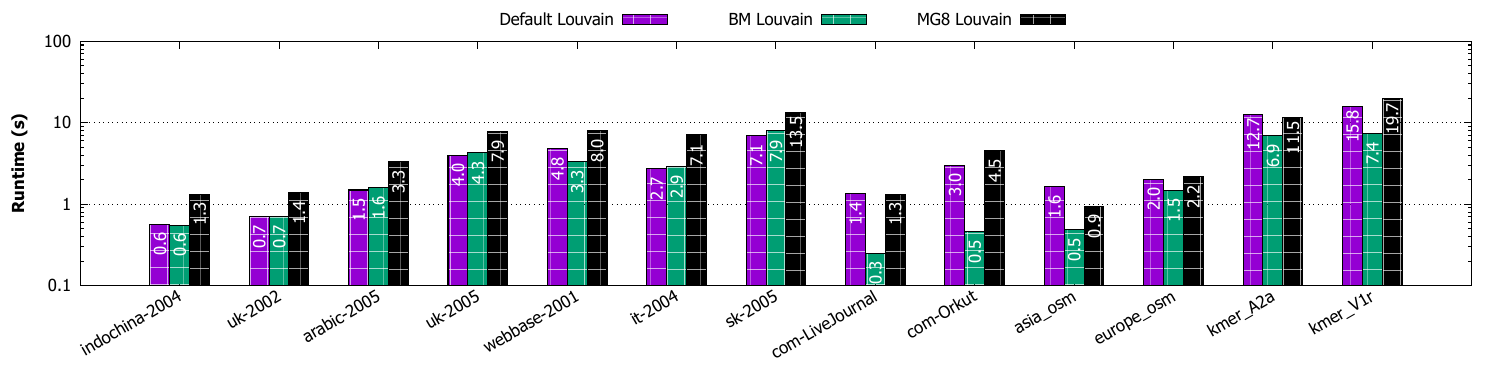}
  } \\[-0ex]
  \subfigure[Modularity of communities obtained with \textit{Default}, \textit{weighted Boyer-Moore (BM) based}, and \textit{weighted Misra-Gries with $8$-slots (MG8) based} Louvain]{
    \label{fig:cmplou--modularity}
    \includegraphics[width=0.98\linewidth]{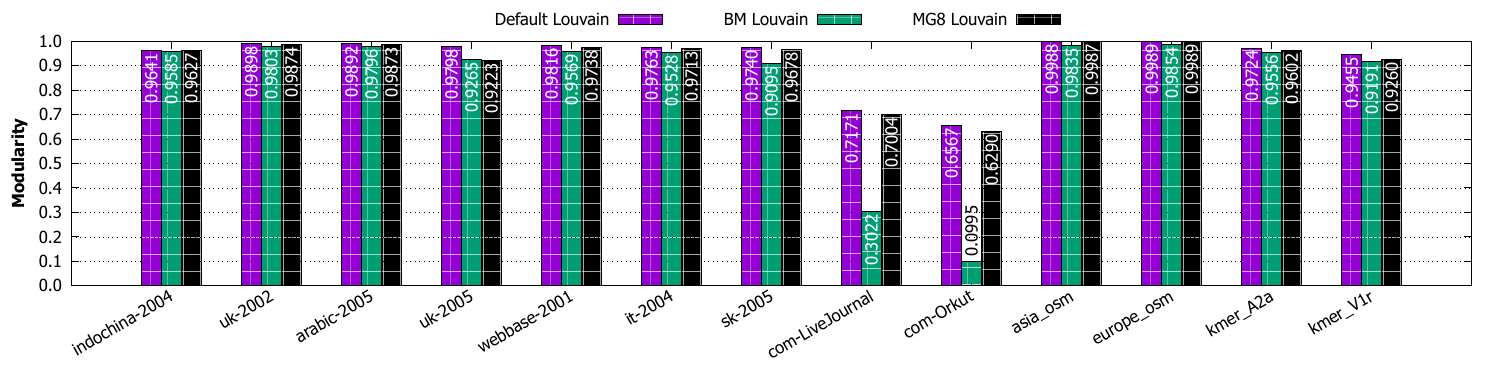}
  } \\[-2ex]
  \caption{Runtime in seconds (log-scale) and Modularity of obtained communities with \textit{Default Louvain}, \textit{weighted Boyer-Moore (BM) based Louvain}, and \textit{weighted Misra-Gries with $8$-slots (MG8) based Louvain} for each graph in the dataset.}
  \label{fig:cmplou}
\end{figure*}

We now evaluate the performance of the parallel implementation of MG8 Louvain (a weighted Misra-Gries based Louvain with $k=8$ slots) and compare it to Default Louvain \cite{sahu2023gvelouvain} and the BM-based Louvain (weighted Boyer-Moore based Louvain) on large graphs, given in Table \ref{tab:dataset}. As outlined in Section \ref{sec:dataset}, we ensure that the graphs are both undirected and weighted. Figure \ref{fig:cmplou} presents the execution times and the modularity of the communities generated by Default, BM, and MG8 Louvain for each graph in the dataset. As shown in Figure \ref{fig:cmplou--runtime}, MG8 Louvain is, on average, $2.07\times$ slower than Default Louvain on web graphs, but demonstrates roughly the same performance on social networks, road networks, and protein k-mer graphs. Regarding modularity, Figure \ref{fig:cmplou--modularity} indicates that MG8 Louvain achieves modularity similar to that of Default Louvain, with the exception of the \textit{uk-2005} web graph. While BM Louvain performs slightly worse in terms of modularity compared to MG8 Louvain on most graphs, it shows a significant drop in modularity on social networks. This discrepancy may arise from the high average degree of these graphs and their lack of a robust community structure. Furthermore, BM Louvain converges much faster on social networks, as illustrated in Figure \ref{fig:cmplou--runtime}, albeit producing lower-quality communities. Consequently, we recommend using BM Louvain for web graphs, road networks, and protein k-mer graphs, but not for social networks. In contrast, MG8 Louvain can be reliably applied to all graph types.

\begin{figure*}[hbtp]
  \centering
  \subfigure[Runtime in seconds (logarithmic scale) with \textit{Default}, \textit{weighted Boyer-Moore (BM) based}, and \textit{weighted Misra-Gries with $64$-slots (MG64) based} Leiden]{
    \label{fig:cmplei--runtime}
    \includegraphics[width=0.98\linewidth]{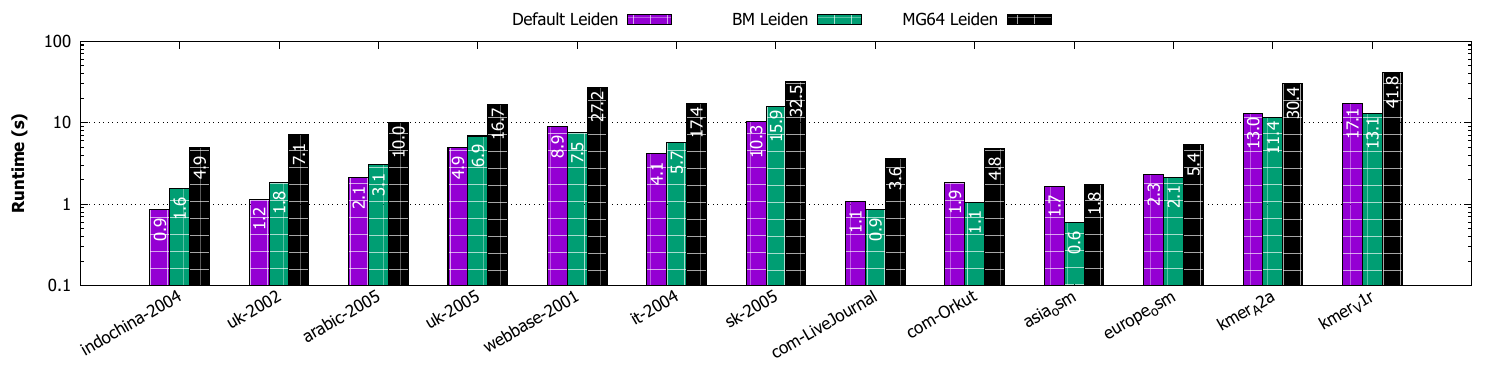}
  } \\[-0ex]
  \subfigure[Modularity of communities obtained with \textit{Default}, \textit{weighted Boyer-Moore (BM) based}, and \textit{weighted Misra-Gries with $64$-slots (MG64) based} Leiden]{
    \label{fig:cmplei--modularity}
    \includegraphics[width=0.98\linewidth]{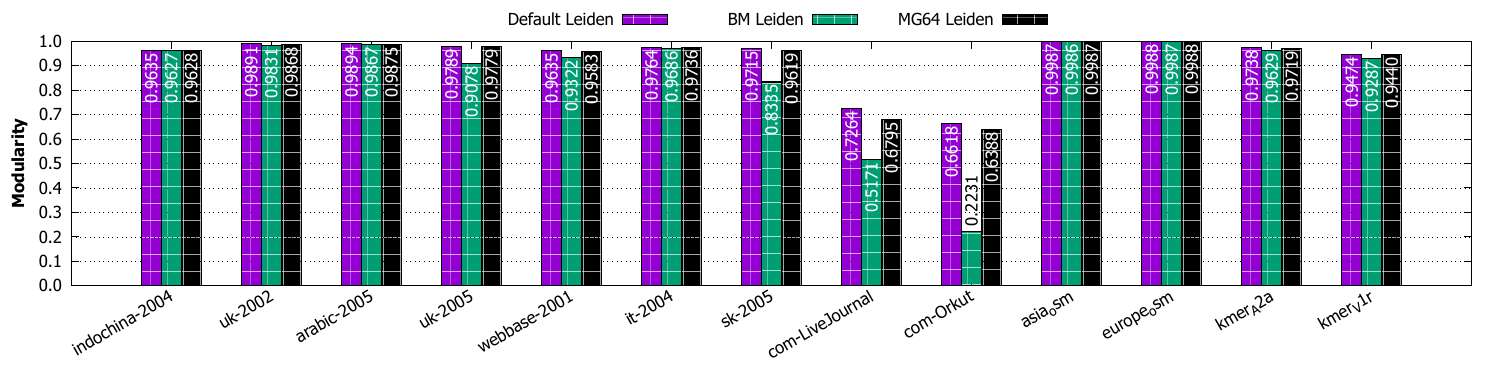}
  } \\[-2ex]
  \caption{Runtime in seconds (log-scale) and Modularity of obtained communities with \textit{Default Leiden}, \textit{weighted Boyer-Moore (BM) based Leiden}, and \textit{weighted Misra-Gries with $64$-slots (MG64) based Leiden} for each graph in the dataset.}
  \label{fig:cmplei}
\end{figure*}

\begin{figure*}[hbtp]
  \centering
  \subfigure[Runtime in seconds (logarithmic scale) with \textit{Default}, \textit{weighted Boyer-Moore (BM) based}, and \textit{weighted Misra-Gries with $8$-slots (MG8) based} LPA]{
    \label{fig:cmprak--runtime}
    \includegraphics[width=0.98\linewidth]{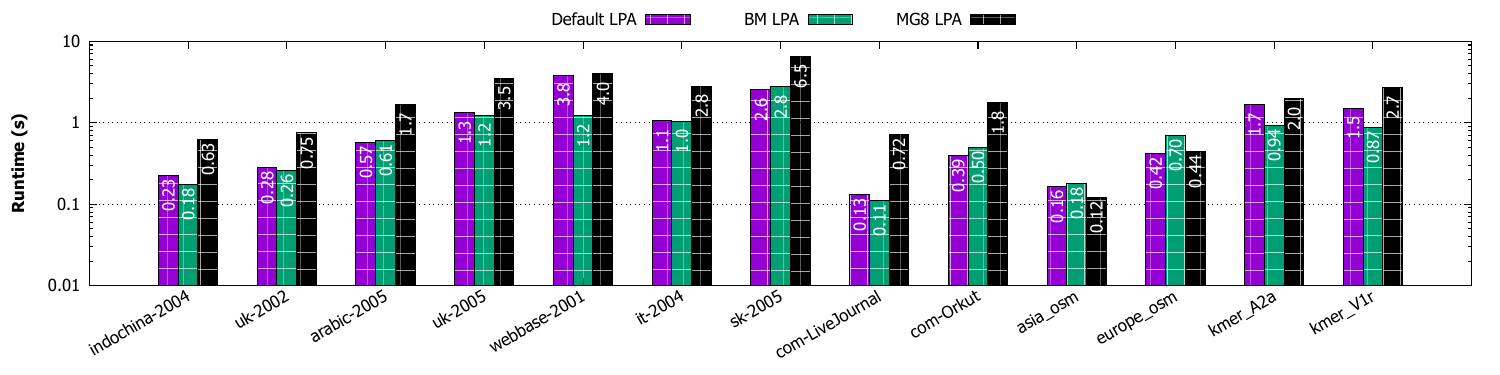}
  } \\[-0ex]
  \subfigure[Modularity of communities obtained with \textit{Default}, \textit{weighted Boyer-Moore (BM) based}, and \textit{weighted Misra-Gries with $8$-slots (MG8) based} LPA]{
    \label{fig:cmprak--modularity}
    \includegraphics[width=0.98\linewidth]{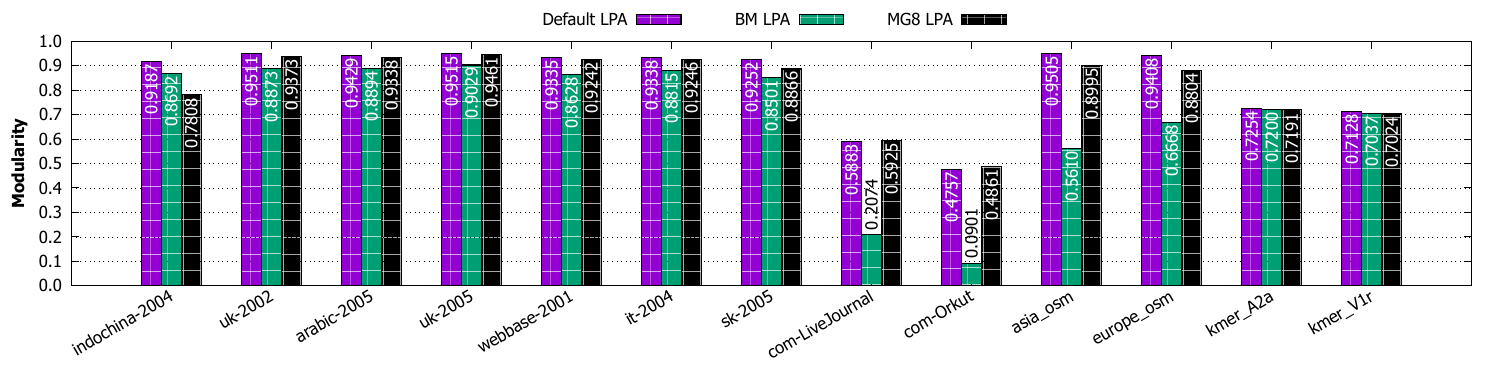}
  } \\[-2ex]
  \caption{Runtime in seconds (log-scale) and Modularity of obtained communities with \textit{Default LPA}, \textit{weighted Boyer-Moore (BM) based LPA}, and \textit{weighted Misra-Gries with $8$-slots (MG8, $2$-scan) based LPA} for each graph in the dataset.}
  \label{fig:cmprak}
\end{figure*}

\subsection{Performance of our MG-based Leiden}
\label{sec:performance-leiden}

Next, we assess the performance of the parallel implementation of MG64 Leiden (MG-based Leiden with $k=64$ slots) and compare it to Default Leiden \cite{sahu2023gveleiden} and BM-based Leiden, across the graphs listed in Table \ref{tab:dataset}. Figure \ref{fig:cmplei} displays the runtime and the modularity of the communities identified by the algorithms for each graph in the dataset. As shown in Figure \ref{fig:cmplei--runtime}, MG64 Leiden is on average $3.15\times$ slower than Default Leiden. In terms of modularity, Figure \ref{fig:cmplei--modularity} indicates that MG64 Leiden achieves a modularity that is only $0.8\%$ lower than Default Leiden. BM Leiden exhibits lower modularity compared to MG64 Leiden, and its performance is particularly poor on social networks and the \textit{sk-2005} web graph. This may be attributed to the high average degree of these graphs, and the weak community structure typical of social networks. BM Leiden also converges significantly faster on social networks, as shown in Figure \ref{fig:cmplei--runtime}, while obtaining these poor quality communities --- but it also converges faster on the \textit{asia\_osm} road network, while obtaining high quality communities. Thus, we recommend employing BM Leiden for road networks and protein k-mer graphs only\ignore{, but not for web graphs or social networks}.\ignore{Similar to MG8 Louvain} However, MG64 Leiden can be reliably utilized across all graph types.

\subsection{Performance of our MG-based LPA}
\label{sec:performance-rak}

Finally, we evaluate the performance of the parallel implementation of MG8 LPA (MG-based LPA with $k=8$ slots) and compare it with Default LPA \cite{sahu2023gvelpa} and BM-based LPA on large graphs (Table \ref{tab:dataset}). Figure \ref{fig:cmprak} illustrates the execution times and the modularity of the communities returned by Default, BM, and MG8 LPA across each graph in the dataset. As depicted in Figure \ref{fig:cmprak--runtime}, MG8 LPA is, on average, $2.78\times$ slower than Default LPA on web graphs and social networks, but only $1.14\times$ slower on road networks and protein k-mer graphs. In terms of modularity, Figure \ref{fig:cmprak--modularity} shows that MG8 LPA achieves a level of modularity comparable to that of Default LPA. While BM LPA performs slightly worse in terms of modularity than MG8 LPA on most graphs, it experiences a notable decrease in modularity on social networks and road networks. Therefore, we recommend using BM LPA for protein k-mer graphs, and possibly web graphs, but not for social networks or road networks. MG8 LPA however can be used reliably applied to all graph types.

\section{Conclusion}
\label{sec:conclusion}
Most existing studies on community detection focus heavily on optimizing performance. However, memory usage becomes a significant challenge, particularly with large graphs on shared-memory systems. In our prior work, we introduced one of the most efficient implementations of the Louvain, Leiden, and Label Propagation Algorithm (LPA). However, these methods incur high memory overhead due to the use of per-thread hashtables, which for a $100$ million vertex graph processed with $64$ threads, can require between $51.2$ GB and $102.4$ GB solely for hashtables. This motivated our exploration of strategies to reduce the memory demands of the algorithms, even if it comes at some cost to performance.

In this work, we proposed to replace the per-thread hashtables in these implementations with a weighted version of the Misra-Gries (MG) sketch. Our experiments showed that MG sketches with $8$, $64$, and $8$ slots for Louvain, Leiden, and LPA, respectively, significantly lower memory usage the memory usage of the implementations --- while suffering only up to $1\%$ decrease in community quality, but with runtime penalties of $1.48\times$, $3.15\times$, and $2.11\times$, respectively. Additionally, we presented a weighted Boyer-Moore (BM) variant for Louvain, Leiden, and LPA, which demonstrated good performance on specific graph types. We believe that these approaches, while introducing some runtime overhead, are well parallelizable and hold potential to surpass current memory-intensive techniques in performance on devices with a large number of threads.

\begin{acks}
I would like to thank Prof. Kishore Kothapalli, Prof. Dip Sankar Banerjee, Josh Bradley\ignore{for identifying the issue of disconnected communities in low-memory Leiden}, and Balavarun Pedapudi for their support.
\end{acks}

\bibliographystyle{ACM-Reference-Format}
\bibliography{main}

\clearpage
\appendix
\section{Appendix}
\subsection{Populating Misra-Gries (MG) sketch}
\label{sec:populate}

The pseudocode for populating a Misra-Gries (MG) sketch, for each neighboring vertex $j$ of a target vertex $i$ in the graph $G$, is shown in Algorithm \ref{alg:populate}. It is used for local-moving, refinement, and aggregation phases of our modified MG Louvain, Leiden, and LPA. The algorithm's inputs include the keys $S_k$ and values $S_v$ in the MG sketch, the current vertex $i$, its neighbor $j$, and the edge weight $w$ connecting them, community membership $C$ for each vertex, and the community bound $C_B$ used during the refinement phase. The parameter $k$ defines the number of slots available in the MG sketch.

To begin, we evaluate two conditions to determine whether to proceed with updating the sketch. First, if the current vertex $i$ is the same as its neighbor $j$ (indicating a self-loop) and self-edges are to be excluded, the function immediately exits (line \ref{alg:populate--exclude-self}) --- self-edges are excluded for the local-moving and refinement phases of the Louvain and Leiden algorithm, and for all iterations of LPA. Next, during the refinement phase (for Leiden algorithm only), if the community bound of vertex $i$ differs from that of vertex $j$, we also exit early, as only connections within the same community bound are relevant (line \ref{alg:populate--check-refine}). If the conditions are met, we attempt to accumulate the edge weight $w$ to the current community $c$ of vertex $i$. We iterate over all slots in the sketch, checking if the community key $S_k[p]$ matches $c$. If so, we add $w$ to the existing weight in $S_v[p]$ (lines \ref{alg:populate--accumulate-begin}-\ref{alg:populate--accumulate-end}). Following this accumulation step, we verify if the community $c$ is already represented in the sketch. A \textit{has} flag is set to track the existence of $c$, and if found, the function terminates early without further modification (lines \ref{alg:populate--exists-begin}-\ref{alg:populate--does-exist}). If the community $c$ does not already exist in the sketch, we search for an empty slot to add it. An empty slot is identified as any position in $S_v$ with a value of $0$. If such a slot is found, we assign the community $c$ to the empty slot in $S_k$ and set the weight $w$ in $S_v$ (lines \ref{alg:populate--empty-begin}-\ref{alg:populate--add-end}). In cases where no empty slots are available, we apply a ``subtractive" operation, reducing each slot's value by $w$, effectively implementing a form of decay for low-frequency communities (lines \ref{alg:populate--subtract-begin}-\ref{alg:populate--subtract-end}). Any slots with a zero value would now be considered empty.

Once the MG sketch has been populated with information from all neighbors of a given vertex $i$, it can be used to select the best community $c^\#$ to which the vertex may move. Algorithm \ref{alg:populate} can also be used to populate the MG sketch with information from the neighbors of all vertices in a community $c$. This can then be used to merge all vertices in $c$ into a single super-vertex, with associated cross-community edges, within the super-vertex graph.

\begin{algorithm}[hbtp]
\caption{Populating Misra-Gries (MG) sketch for each neighboring vertex of a given vertex in the graph (Louvain, Leiden, LPA).}
\label{alg:populate}
\begin{algorithmic}[1]
\Require{$S_k, S_v$: Keys, values array of the MG sketch}
\Require{$i, j, w$: Current vertex, its neighbor, and associated edge weight}
\Require{$C$: Community membership of each vertex}
\Require{$C_B$: Community bound of each vertex (refine phase only)}
\Ensure{$k$: Number of slots in the MG sketch}

\Statex

\Function{populateSketch}{$S_k, S_v, i, j, w, C, C_B$} \label{alg:populate--begin}
  \If{\textbf{exclude self} $= 1$ \textbf{and} $i = j$} \ReturnInline{} \label{alg:populate--exclude-self}
  \EndIf
  \If{\textbf{is refine phase and} $C_B[i] \neq C_B[j]$} \ReturnInline{} \label{alg:populate--check-refine}
  \EndIf
  \State $c \gets C[i]$
  \State $\rhd$ Add edge weight to community
  \ForAll{$p \in [0, k)$} \label{alg:populate--accumulate-begin}
    \State $S_v[p] \gets S_v[p] + w$ \textbf{if} $S_k[p] = c$ \textbf{else} $S_v[p]$
  \EndFor \label{alg:populate--accumulate-end}
  \State $\rhd$ Check if community is already in the list
  \State $has \gets 0$ \label{alg:populate--exists-begin}
  \ForAll{$p \in [0, k)$}
    \State $has \gets has \vee -1$ \textbf{if} $S_k[p] = c$ \textbf{else} $has$
  \EndFor \label{alg:populate--exists-end}
  \If{$has \neq 0$} \ReturnInline{} \label{alg:populate--does-exist}
  \EndIf
  \State $\rhd$ Find empty slot
  \State $e \gets -1$ \label{alg:populate--empty-begin}
  \ForAll{$p \in [0, k)$}
    \If{$S_v[p] = 0$} $e \gets p$
    \EndIf
  \EndFor \label{alg:populate--empty-end}
  \If{$e \geq 0$}
    \State $S_k[e] \gets c$ \label{alg:populate--add-begin}
    \State $S_v[e] \gets w$ \label{alg:populate--add-end}
  \Else
    \ForAll{$p \in [0, k)$} \label{alg:populate--subtract-begin}
      \State $S_v[p] \gets max(S_v[p] - w, 0)$
    \EndFor \label{alg:populate--subtract-end}
  \EndIf
\EndFunction \label{alg:populate--end}
\end{algorithmic}
\end{algorithm}

\subsection{Using smaller hashtables}
\label{sec:small-hashtables}

To minimize memory usage of per-thread hashtables, used in the Louvain, Leiden, and LPA algorithms, we first attempt to scale down the size of the \textit{values} array, and use a mod-based lookup instead of direct indexing. Collisions can occur in such a hashtable. When a collision occurs, only the first key is added to the keys list, while values of colliding keys accumulated into the same slot. While this approach sacrifices the ability to distinguish between colliding keys, it seems to be effective for heuristic community detection methods like Louvain, Leiden, and LPA. In fact, reducing the size of the values array by $3000\times$ maintains community quality, and even offers slightly faster runtimes.

\begin{figure*}[hbtp]
  \centering
  \subfigure[Overall result]{
    \label{fig:smlei-runtime--mean}
    \includegraphics[width=0.38\linewidth]{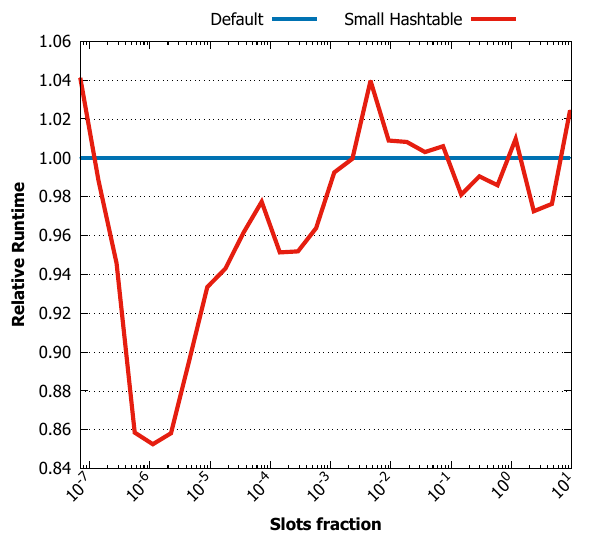}
  }
  \subfigure[Results on each graph]{
    \label{fig:smlei-runtime--all}
    \includegraphics[width=0.58\linewidth]{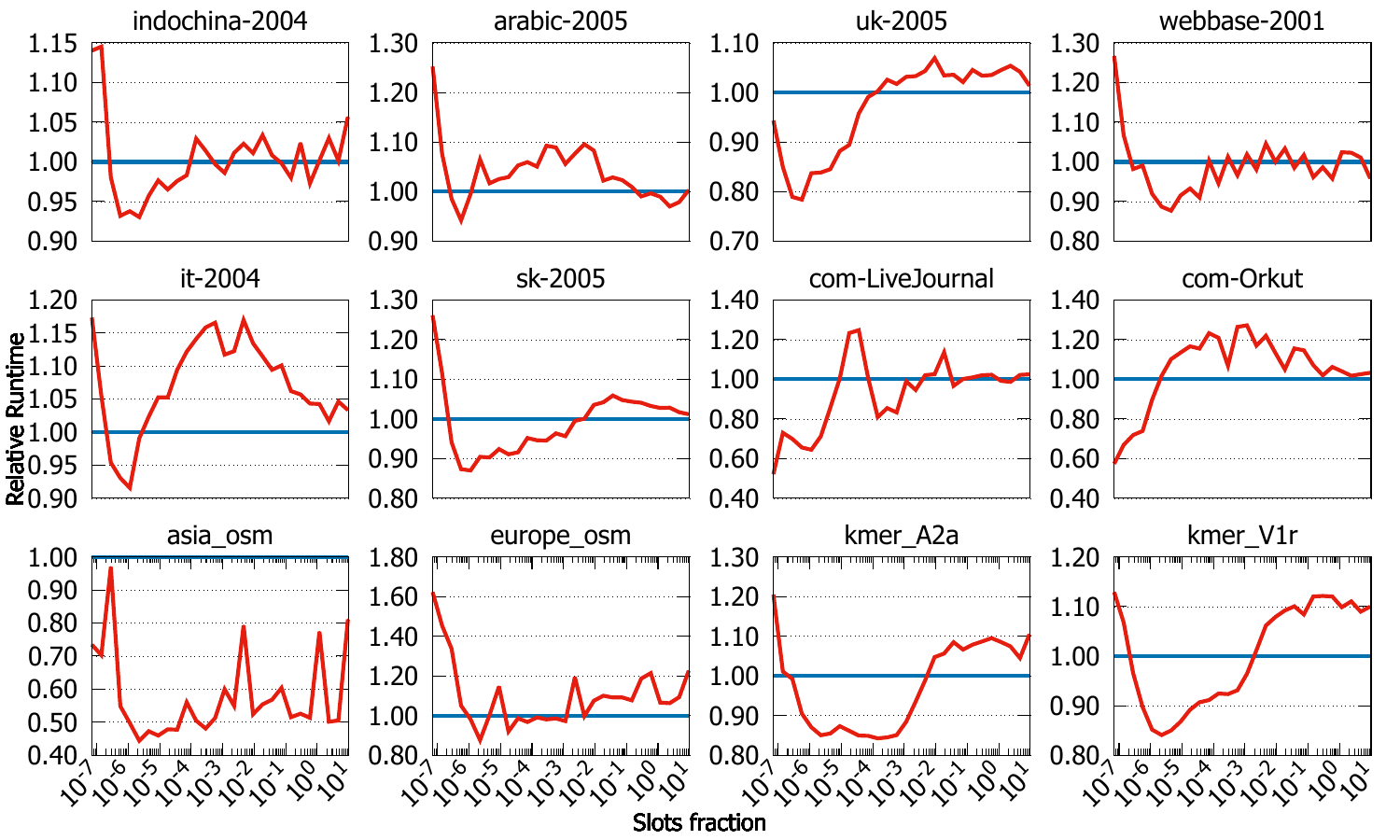}
  } \\[-2ex]
  \caption{Relative runtime of our parallel small-hastable based Leiden, where only old keys with the same hash are inserted, while values of colliding keys accumulated into the same slot --- compared to Default Leiden \cite{sahu2023gveleiden} which utilizes a keys list with as full size values array as its per-thread hashtable.}
  \label{fig:smlei-runtime}
\end{figure*}

\begin{figure*}[hbtp]
  \centering
  \subfigure[Overall result]{
    \label{fig:smlei-modularity--mean}
    \includegraphics[width=0.38\linewidth]{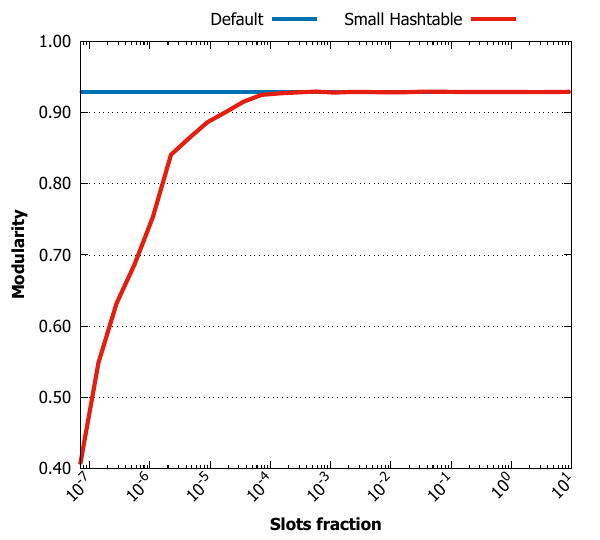}
  }
  \subfigure[Results on each graph]{
    \label{fig:smlei-modularity--all}
    \includegraphics[width=0.58\linewidth]{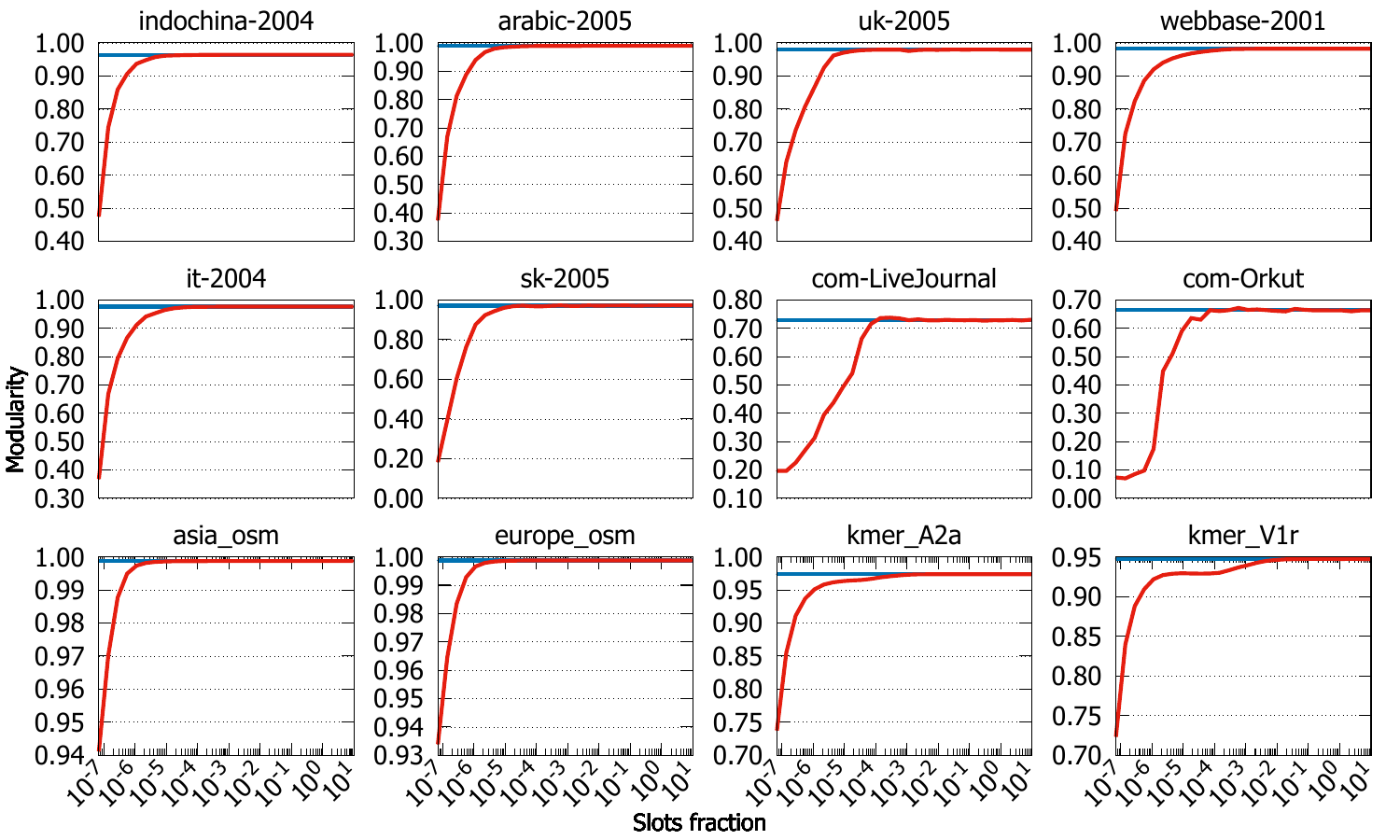}
  } \\[-2ex]
  \caption{Relative modularity of our parallel small-hastable based Leiden, where only old keys with the same hash are inserted, while values of colliding keys accumulated into the same slot --- compared to Default Leiden \cite{sahu2023gveleiden} which utilizes a keys list with as full size values array as its per-thread hashtable.}
  \label{fig:smlei-modularity}
\end{figure*}

In our experiments, we vary the hashtable slots from $2$ to $2^{28}$ and measure the performance of small-hashtable based Leiden. Then, for each graph in Table \ref{tab:dataset}, we observe the impact of the ``slots fraction” (the number of slots in the small per-thread hashtable, relative to the number of vertices in the graph $|V|$) on runtime and modularity. As shown in Figure \ref{fig:smlei-runtime}, reducing the slots fraction initially improves runtime due to locality benefits. However, past a certain point, runtime increases as the algorithm takes more iterations to compensate for degraded community quality. This degradation in quality is a consequence of having fewer slots to differentiate communities, leading to more iterations for convergence. In fact, this can clearly be seen in Figure \ref{fig:smlei-modularity}, where the trend is a smooth curve showing how the reducing hashtable size (in terms of slots fraction) reduces the quality of obtained communities. The optimal balance appears around a slots fraction of $3 \times 10^{-4}$, where runtime somewhat improves and memory use is significantly reduced (by around $3000\times$). Using smaller hashtables thus cuts memory usage, albeit without major runtime gains. These results are encouraging, and we plan further reductions in the algorithm’s memory footprint, which are discussed in Section \ref{sec:approach}.

\begin{figure*}[!hbt]
  \centering
  \subfigure[Source code of our MG8 kernel]{
    \label{fig:about-mg8code--source}
    \raisebox{3ex}{\includegraphics[width=0.49\linewidth]{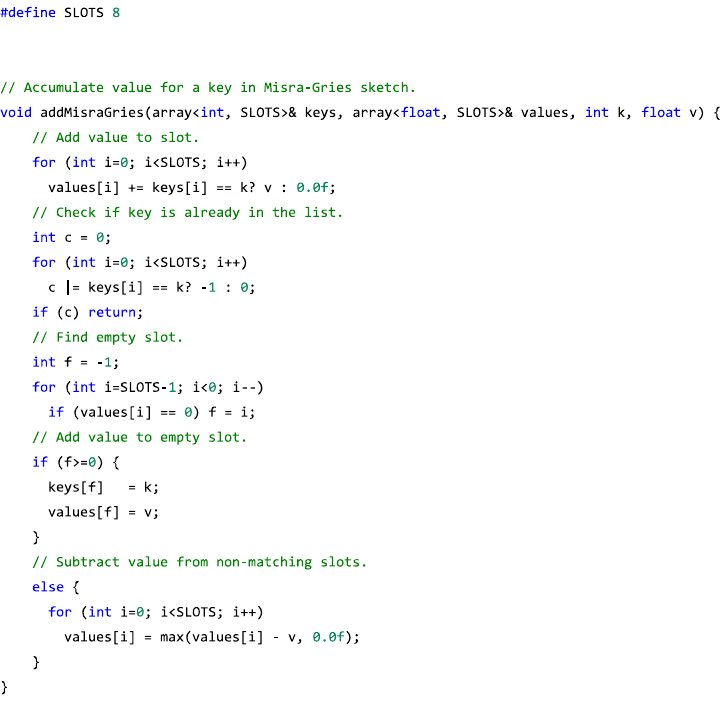}}
  }
  \subfigure[Translated machine code of our MG8 kernel]{
    \label{fig:about-mg8code--assembly}
    \includegraphics[width=0.39\linewidth]{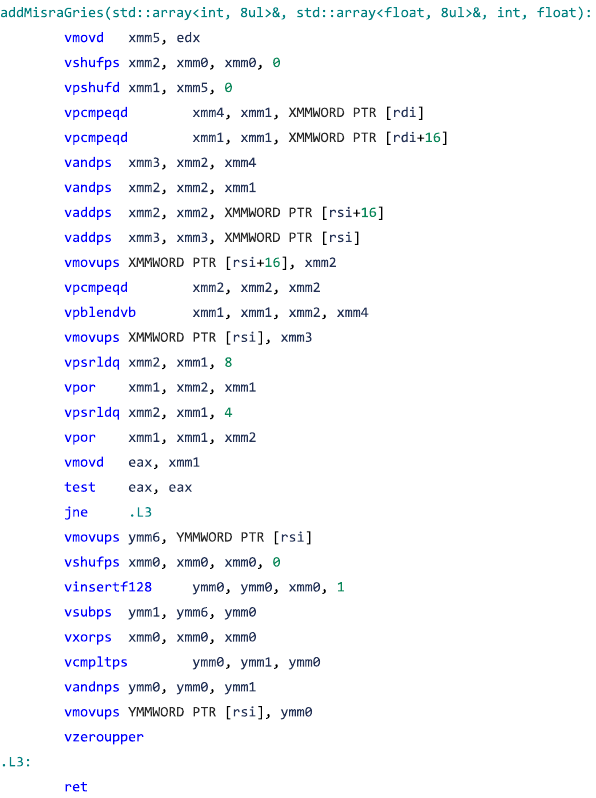}
  } \\[-2ex]
  \caption{The above figure shows our Misra-Gries with $8$-slots (MG8) kernel for accumulating the community (key) with associated weight (value) to the Misra-Gries sketch --- used with Leiden, Louvain, and LPA algorithms --- along with its compiled x86 machine code using GCC 13 with \texttt{-O3} and \texttt{-mavx} flags. As the figure shows, the code, which is run multiples times for each edge in the graph (and super-vertex graphs), can be efficiently translated by the compiler into vector machine instructions with minimal conditional jumps. For this to happen, it is important to write the source code as a fairly straightforward series of for-loops of a fixed width (using compile-time constants) --- if the loops are not straightforward, the compiler may not be able to deduce the suitable vector instructions to use, and instead emit non-vector code with loop unrolling --- this may not have good performance. We use the Compiler Explorer, by Matt Godblot, for this interactive code translation.}
  \label{fig:about-mg8code}
\end{figure*}

\end{document}